\documentclass[3p,times,procedia]{elsarticle}
\usepackage{ecrc}
\usepackage[bookmarks=false]{hyperref}
    \hypersetup{colorlinks,
      linkcolor=blue,
      citecolor=blue,
      urlcolor=blue}

\usepackage{tikz}
\usetikzlibrary{shapes.geometric}
\usepackage{booktabs}
\usepackage{array}
\usepackage{tabularx}
\usepackage{amssymb}   %

\volume{00}
\firstpage{1}
\journalname{Procedia Computer Science}
\runauth{Rudd et al.}
\jid{procs}

\begin{document}
\begin{frontmatter}

\dochead{30th International Conference on Knowledge-Based and Intelligent
Information \& Engineering Systems (KES 2026)}

\title{Architecting the Corporate AI Infrastructure: The Evolving Roles of AI
Governance Committees and Chief Ethics Officers}

\author[a]{David Hason Rudd\corref{cor1}}
\ead{david.hasonrudd@uts.edu.au}

\author[a]{Dennis Singh}

\author[b]{Cesar Sanin}

\author[c]{Bryn James}

\address[a]{University of Technology Sydney, Australia}
\address[b]{University of New England, Australia}
\address[c]{Australian Institute of Higher Education}

\cortext[cor1]{Corresponding author.}

\begin{abstract}
The accelerating integration of artificial intelligence~(AI) into core
business operations has created an urgent need for robust corporate governance
infrastructures capable of managing complex ethical, legal, and operational
risks. While organizations have increasingly adopted high-level AI ethics
principles, a critical implementation gap persists: fewer than half of
organizations that cite AI governance as a boardroom priority have a
formalized governance framework in place. This paper addresses that gap by
proposing a structured corporate AI governance architecture (CAIGA) centred on two mutually reinforcing pillars. First, it examines the evolving mandate of specialized C-suite roles, specifically Chief Privacy Officers~(CPOs) and Chief Ethics Officers~(ChEOs), in formalizing accountability for AI usage across the enterprise. Second, it argues that cross-functional AI Governance Committees, uniting representatives from legal, technology, data science, cybersecurity, and business units, are the essential organizational mechanism for translating ethical principles into operational decisions, including strategic go/no-go approvals for new AI initiatives. Drawing on a conceptual review of peer-reviewed governance literature, regulatory frameworks including the EU AI Act, and documented industry practice, this paper contributes a replicable governance architecture that embeds clear accountability policies, structured risk management workflows, and an organization-wide ethical AI culture. The framework is illustrated through
application to the financial services sector. The findings have direct
implications for corporate leaders, risk professionals, and policymakers
navigating the rapidly evolving global AI regulatory landscape.
\end{abstract}

\begin{keyword}
AI governance \sep AI ethics committees \sep Chief Ethics Officer \sep
Chief Privacy Officer \sep corporate AI infrastructure \sep risk management
\sep responsible AI \sep EU AI Act \sep go/no-go decisions \sep ethical AI
culture \sep financial services
\end{keyword}

\end{frontmatter}

\section{Introduction}
\label{sec:intro}

Artificial intelligence is no longer an emerging technology on the periphery
of corporate strategy; it is now embedded in hiring decisions, credit
assessments, supply chain management, customer interactions, and clinical
diagnostics. This pervasive integration brings transformative opportunity, but
also a governance crisis that organizations are ill-equipped to address. A 2024
EY Pulse Survey found that while 95~percent of senior leaders report their
organizations are actively investing in AI, only 34~percent have incorporated
AI governance frameworks, and just 32~percent are addressing bias in AI
models~\cite{EY2024}. Compounding this, the AI Incident Database recorded a
26~percent rise in reported AI ethical incidents from 2022 to 2023, with
credible indicators of a further 32~percent increase in 2024~\cite{AIID2024}.
As Hickman and Petrin~\cite{Hickman2021} note, corporate boards are increasingly
facing direct liability for algorithmic failures under evolving company law,
necessitating a paradigm shift in how AI is governed at the enterprise level.
The gap between AI adoption velocity and governance maturity is not merely an
organizational inefficiency; it is a source of escalating reputational,
legal, and operational risk~\cite{Tandfonline2025, NACDO2024}.

Existing governance literature acknowledges this challenge but predominantly
addresses it through one of two lenses: either proposing broad ethical
principles and policy frameworks at the regulatory level, or examining
technical mitigation strategies such as algorithmic auditing and bias
detection~\cite{Correa2023, ribeiro2025toward}. What remains underexplored is
the organizational and structural layer that sits between these two
levels, specifically how corporations should architect their internal
governance infrastructure to make ethical AI governance repeatable,
accountable, and scalable in practice~\cite{Birkstedt2023}. Questions such as
who is accountable when an AI system produces a biased or harmful decision,
how cross-functional teams should coordinate on go/no-go approvals for new AI
deployments, and what role specialized executive officers should play in
embedding ethical norms across business units remain insufficiently answered
in the literature~\cite{Mantymaki2022}.

This paper addresses that structural gap by proposing a comprehensive corporate
AI governance architecture organized around two interdependent pillars. The
first concerns specialized leadership: the roles of CPOs and ChEOs are evolving
beyond compliance gatekeeping toward proactive, enterprise-wide stewardship of
AI ethics, requiring new mandates, reporting structures, and operational
authority. The second concerns organizational design: cross-functional AI
Governance Committees, drawing membership from legal, cybersecurity,
technology, data science, business, and audit and risk functions, are
positioned as the essential deliberative mechanism for overseeing AI systems
and making structured, auditable decisions on new AI initiatives.

The paper makes three specific contributions to the literature. First, it
synthesizes current evidence on the evolving mandate of CPOs and ChEOs in the
context of enterprise AI governance. Second, it articulates the composition,
decision-making processes, and lifecycle role of cross-functional AI Governance
Committees, grounded in empirical research and industry practice. Third, it
proposes a replicable governance architecture integrating these two pillars
with risk management frameworks and ethical AI culture, illustrated through
application to financial services, and aligns the framework with the EU AI
Act's risk-tiered classification system.

The remainder of the paper is organized as follows. Section~\ref{sec:related} reviews related 
work. Section~\ref{sec:methodology} describes the research methodology and presents the proposed 
governance architecture. Section~\ref{sec:discussion_conclusion} discusses regulatory alignment, addresses 
the limitations of the framework, and concludes the paper.

\section{Related Work}
\label{sec:related}

\subsection{AI Governance: Imperatives and the Implementation Gap}

The rapid integration of artificial intelligence into business operations has
generated mounting pressure on organizations to manage complex ethical, legal,
and operational risks~\cite{Camilleri2024}. A meta-analysis of 200 global AI
governance guidelines found that despite widespread normative consensus on
principles such as fairness, transparency, and accountability, concrete
implementation mechanisms remain inconsistently applied across organizations
and jurisdictions~\cite{Correa2023}. Scholars have consistently observed that
the gap between the design of AI systems and our understanding of their
ethical implications can carry severe consequences for individuals, groups,
and whole societies~\cite{Sison2023}.

A systematic literature review by Birkstedt et al.~\cite{Birkstedt2023}
confirms that AI governance is a rapidly growing yet fragmented research area,
and that while organizations have produced extensive documentation of ethical
principles, operationalizing these principles into repeatable organizational
processes remains a nascent and underserved area of scholarship.
M\"{a}ntym\"{a}ki et al.~\cite{Mantymaki2022} elaborate on this implementation
gap through their Hourglass Model, demonstrating that ethical principles must
pass through multiple organizational filters (policy, process, and
culture) before manifesting as consistent governance behaviour. Translating
AI ethics principles into practice therefore requires more than aspirational
statements; it demands clearly defined accountability structures, formal review
processes, and embedded organizational routines~\cite{Madanchian2025}. To
bridge this gap, M\"{o}kander and Floridi~\cite{Mokander2021} propose
ethics-based auditing as a critical mechanism to evaluate automated
decision-making systems. However, they caution that audits alone are
insufficient without a broader corporate governance apparatus to enforce their
findings, further emphasizing the need for structural accountability.

\subsection{Governance Structures: Leadership Roles and Cross-Functional
Committees}

The literature increasingly identifies specialized C-suite leadership as a
critical mechanism for institutionalizing AI governance responsibilities.
Camilleri~\cite{Camilleri2024} identifies the promotion of accountability and
transparency as core dimensions of AI governance, arguing that organizations
must assign named executive responsibility for these dimensions rather than
treating them as collective, diffuse obligations. M\"{a}ntym\"{a}ki et
al.~\cite{Mantymaki2022} further argue that effective AI governance requires
formal ownership at the executive level, with governance functions embedded
in defined roles rather than absorbed into existing compliance or legal
responsibilities. Empirical survey data from the International Association of
Privacy Professionals (IAPP) underscores this trend: 69~percent of surveyed
Chief Privacy Officers reported acquiring additional formal responsibility for
AI governance in 2024, confirming the CPO's growing role as a primary
institutional anchor for enterprise AI ethics~\cite{IAPP2024}.

Complementing individual leadership roles, the establishment of cross-functional
AI Governance Committees has emerged as an empirically supported best
practice. Hadley et al.~\cite{Hadley2024} conducted the first systematic
empirical study of Algorithm Review Boards (ARBs) across organizations,
finding that effective boards require a mix of technical and non-technical
experts spanning privacy, ethics, legal, and sociotechnical expertise, and
that their design should be tailored to specific sectoral and organizational
contexts. Their findings showed that health and financial services sectors
maintained the most robust review structures, providing a validated model for
the design of AI Governance Committees in regulated industries.

Birkstedt et al.~\cite{Birkstedt2023} further identify cross-functional
governance structures as one of the most cited organizational mechanisms in
the AI governance literature, while also noting that most existing frameworks
remain underspecified with respect to the committee's precise composition,
decision authority, and interaction with executive leadership. The literature
further emphasizes that siloed departmental functions are insufficient for
managing AI risk at enterprise scale: AI governance requires defined workflows
and collaborative decision-making mechanisms spanning the full AI lifecycle,
from data ingestion and model training to deployment and post-market
monitoring~\cite{Mantymaki2022}. Cybersecurity is particularly identified as
a dimension that must be integral to any AI governance structure, given that
AI systems introduce novel attack surfaces including prompt injection, model
inversion, and adversarial manipulation~\cite{Camilleri2024}.

\subsection{Ethical AI Culture and Competitive Advantage}

Beyond formal structures, the literature emphasizes the need for an
organization-wide ethical AI culture as a necessary complement to structural
governance mechanisms~\cite{Sison2023}. Madanchian and
Taherdoost~\cite{Madanchian2025} argue that responsible AI adoption requires
not only technical and procedural controls but also a supporting organizational
culture in which ethical considerations are embedded in decision-making at all
levels. This cultural dimension is closely linked to competitive outcomes:
organizations that treat AI ethics as a core competency, one that integrates
legal compliance, stakeholder trust, and brand reputation, systematically
outperform peers in building sustainable AI programs~\cite{Correa2023}.

Corporate governance scholars further argue that translating ethical principles
into practice demands clear accountability policies, dedicated risk management
frameworks, and sustained board-level commitment to fostering an ethical AI
culture~\cite{Camilleri2024}. The EU AI Act's risk-based classification system
has catalysed this process by creating regulatory consequences for ethical
failures, thereby transforming the business case for proactive governance from
reputational to legal~\cite{EUAIAct2024}.

\subsection{Research Gap and Contribution}

Despite the growing body of governance literature, a persistent gap remains.
As Birkstedt et al.~\cite{Birkstedt2023} conclude, existing research has
documented what AI governance should achieve but has provided limited guidance
on how organizations should architect the internal structures, including roles,
committees, workflows, and escalation processes, through which governance is
operationalized. Accountability gaps persist particularly in multi-stakeholder
AI ecosystems, where responsibility for ethical lapses across vendors, cloud
providers, and internal teams remains ambiguous~\cite{Sison2023}. The
practical mechanics of how cross-functional committees and executive officers
should interact during strategic go/no-go decisions, and how those decisions
should be documented, escalated, and audited, remain underexplored. This
paper bridges that gap by proposing a governance architecture that
operationalizes these roles, decision workflows, and accountability mechanisms
in a structured and repeatable way aligned with current regulatory frameworks.

\section{Methodology}
\label{sec:methodology}

This paper adopts a conceptual review methodology to develop a governance
framework~\cite{Madanchian2025}. The review was guided by three core research
inquiries: (RQ1)~identifying the most effective organizational structures for
institutionalizing AI governance responsibilities at the executive level;
(RQ2)~determining how cross-functional AI Governance Committees should be composed and
empowered to enable structured go/no-go decision-making; and (RQ3)~exploring how
corporate AI governance frameworks can align with emerging regulatory requirements,
specifically the EU AI Act.

A systematic search was conducted across Scopus, Web of Science, and Google
Scholar using the Boolean query strings: (AI governance OR
artificial intelligence governance) AND (corporate OR enterprise
OR organisation) AND (AI ethics committee OR governance
committee OR algorithm review board) AND (accountability OR
decision-making). The search was bounded to publications from January
2019 to April 2025, reflecting the period of intensifying regulatory and
scholarly activity following the OECD AI Principles (2019)~\cite{OECD2019}. 
An initial pool of 187 candidate sources was identified; after title and 
abstract screening for relevance to organizational governance (as opposed to 
technical AI safety alone), 42 sources were selected for full-text review, 
of which 20 peer-reviewed articles, regulatory documents, and authoritative 
industry reports met all inclusion criteria and are cited in this paper.

Inclusion criteria required sources to address at least one of: corporate AI
governance structures, C-suite accountability for AI, cross-functional
committee design, or risk-based AI classification frameworks. Sources focusing
exclusively on technical AI safety without organizational governance dimensions
were excluded. Thematic synthesis was applied to identify convergent findings
and persistent gaps in the literature, which the proposed framework is designed
to address.

\subsection{Proposed Governance Framework}
\label{sec:framework}

\subsubsection{Architecture Overview}

\begin{figure}[h]
\centering
\begin{tikzpicture}[
  >=stealth,
  font=\small,
  tierbox/.style = {draw=black, thick, rounded corners=2pt, inner sep=2ex},
  gate/.style    = {diamond, draw=black, thick, fill=yellow!15, aspect=2.5,
                    inner sep=2pt, font=\footnotesize\bfseries},
]
\node[tierbox, fill=blue!15] (board) at (0, 0) {
  \begin{tabular}{c}
    \textbf{Tier 1: Board of Directors} \\[1ex]
    AI Risk Sub-Committee $\cdot$ Strategic \\
    Oversight $\cdot$ Final Escalation Authority
  \end{tabular}};
\node[tierbox, fill=teal!15] (cpo) at (-2.6, -2.4) {
  \begin{tabular}{c}
    \textbf{CPO} \\[0.5ex] Chief Privacy Officer
  \end{tabular}};
\node[tierbox, fill=teal!15] (cheo) at (2.6, -2.4) {
  \begin{tabular}{c}
    \textbf{ChEO / CAIO} \\[0.5ex] Chief Ethics \\ / AI Officer
  \end{tabular}};
\node[tierbox, fill=orange!15] (aigc) at (0, -4.8) {
  \begin{tabular}{c}
    \textbf{Tier 3: Cross-Functional AI Governance Committee (AIGC)} \\[1.5ex]
    Legal $\cdot$ Cybersecurity $\cdot$ Data Science $\cdot$ \\
    Technology $\cdot$ Business $\cdot$ Audit \& Risk
  \end{tabular}};
\node[gate] (gate) at (0, -7.2) {Go / No-Go Gate};
\node[tierbox, fill=green!15] (bu1) at (-3.5, -9)
      {\begin{tabular}{c} Business \\ Unit A \end{tabular}};
\node[tierbox, fill=green!15] (bu2) at (0, -9)
      {\begin{tabular}{c} Business \\ Unit B \end{tabular}};
\node[tierbox, fill=green!15] (bu3) at (3.5, -9)
      {\begin{tabular}{c} Business \\ Unit C \end{tabular}};
\draw[->, thick] (board.south) -- (0, -1.2) -| (cpo.north);
\draw[->, thick] (board.south) -- (0, -1.2) -| (cheo.north);
\draw[thick, dashed] (cpo.south) -- (-2.6, -3.5) -- (2.6, -3.5) -- (cheo.south);
\draw[->, thick, dashed] (0, -3.5) -- (aigc.north);
\draw[->, thick] (aigc.south) -- (gate.north);
\draw[->, thick] (gate.south) -- (0, -8.1) -| (bu1.north);
\draw[->, thick] (gate.south) -- (bu2.north);
\draw[->, thick] (gate.south) -- (0, -8.1) -| (bu3.north);
\draw[->, thick, dashed] (bu1.west) -- (-5.2, -9) --
  node[left, align=right] {\footnotesize Audit / \\ \footnotesize Incident}
  (-5.2, -4.8) -- (aigc.west);
\end{tikzpicture}
\caption{Corporate AI Governance Architecture (CAIGA): three-tier authority
structure with embedded Go/No-Go decision gate. Shading denotes tier:
Board (blue), Executive Leadership (teal), AIGC (orange), Business Units
(green), Go/No-Go Gate (amber).}
\label{fig:governance}
\end{figure}
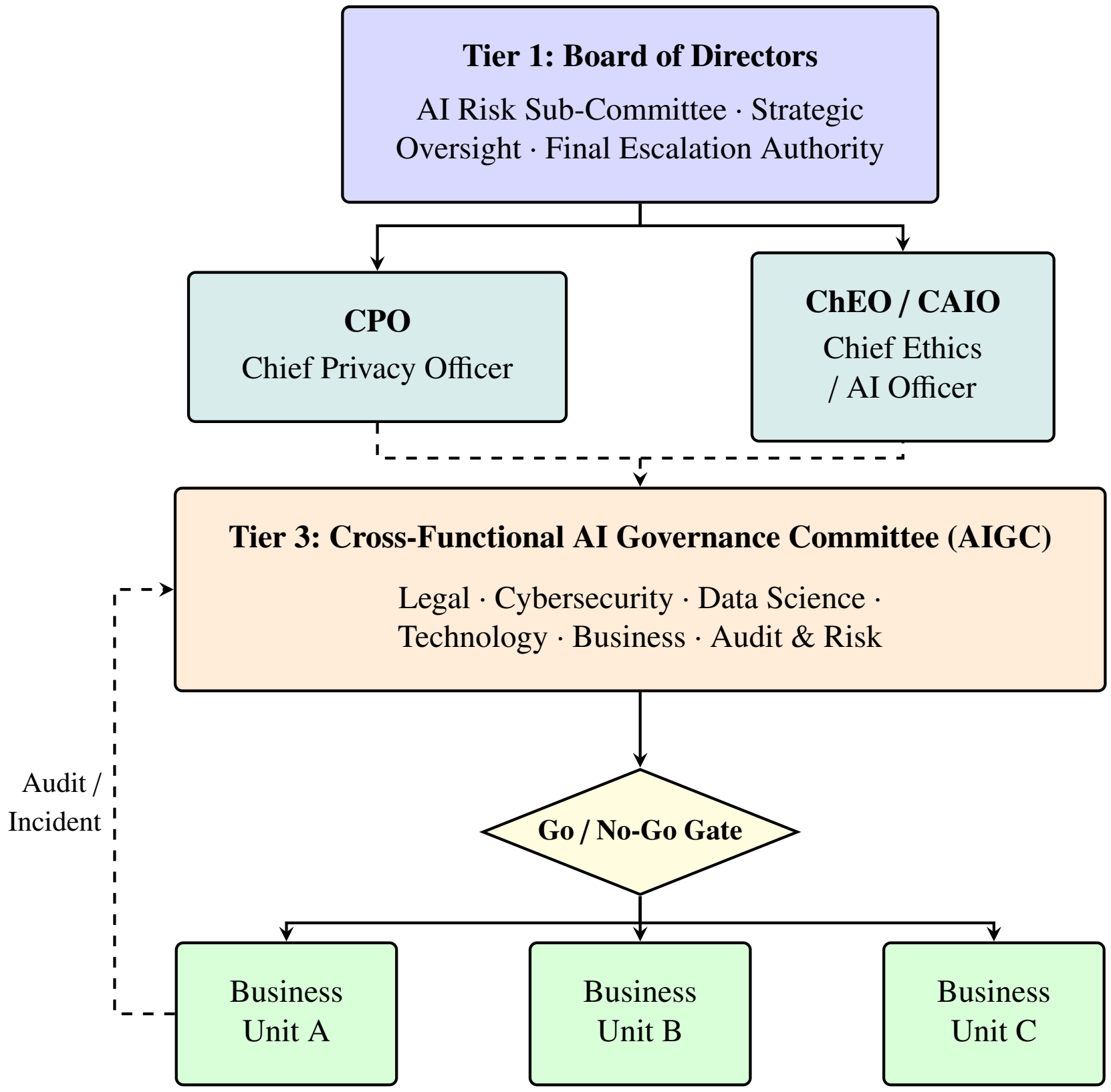

The proposed Corporate AI Governance Architecture (CAIGA) rests on three
interdependent tiers of authority and accountability: (i)~Board-Level
Oversight, (ii)~Executive AI Leadership, and (iii)~the Cross-Functional AI
Governance Committee (AIGC). These tiers operate through structured
information flows, defined escalation paths, and a formal go/no-go decision
process embedded into the AI initiative lifecycle. Figure~\ref{fig:governance}
illustrates the architecture.

\noindent Tier 1: Board of Directors.
The board maintains ultimate accountability for AI risk through a dedicated
AI Risk Sub-Committee. It sets the organization's AI risk appetite, approves
enterprise-wide AI governance policies, and serves as the final escalation
authority for high-stakes or unresolved decisions from the AIGC.

\noindent Tier 2: Executive AI Leadership.
The CPO and ChEO/CAIO serve as the dual executive custodians of the governance
framework. The CPO governs data privacy, consent, and compliance with data
protection regulations. The ChEO/CAIO owns the ethical framework, monitors AI
incident trends, chairs the AIGC, and maintains the board-level reporting
relationship. Both officers hold joint authority to trigger emergency review of
any AI initiative and veto power subject to board ratification for high-risk
deployments.

\noindent Tier 3: AI Governance Committee (AIGC).
The AIGC is the operational heart of the framework: a standing, cross-functional
body comprising representatives from Legal, Cybersecurity, Data Science,
Technology, Business, and Audit \& Risk, consistent with the empirically
validated ARB design identified by Hadley et al.~\cite{Hadley2024}.

\subsubsection{The Go/No-Go Decision Process}
\label{ssec:gonogo}

Any business unit or technology team seeking to deploy a new AI system or
materially modify an existing one must submit an AI Initiative Proposal~(AIP)
to the AIGC. The review process follows five sequential stages, beginning when
the business unit submits an AIP that includes the use case description,
intended data sources, affected stakeholder groups, anticipated business benefit,
and a preliminary self-assessment of the risk tier. Following this submission,
the AIGC Secretariat (Audit \& Risk function) conducts a risk triage by assigning
a formal risk classification within five business days, applying the criteria
outlined in Table~\ref{tab:gonogo}. During the subsequent committee review, each
AIGC member evaluates the AIP within their respective domain within ten business
days. Specifically, Legal assesses regulatory exposure, Cybersecurity assesses
system integrity and attack surface, Data Science evaluates model fairness and
data quality, Business determines commercial impact, and Audit \& Risk verifies
documentation completeness. Based on these evaluations, the AIGC renders a formal
decision (Approved, Approved with Conditions, Deferred for Remediation, or Rejected),
which is documented with an explicit rationale and signed by the Chair. Finally,
approved initiatives transition to post-deployment monitoring, where they are
subject to a defined monitoring schedule with mandatory incident reporting, and
any material changes to the system trigger a mandatory re-review.

\begin{table}[!ht]
\caption{AI Initiative Go/No-Go Decision Matrix}
\label{tab:gonogo}
\begin{tabularx}{\linewidth}{@{} l >{\raggedright\arraybackslash}X
  >{\raggedright\arraybackslash}X >{\raggedright\arraybackslash}X @{}}
\toprule
\textbf{Risk Tier} & \textbf{Trigger Criteria} & \textbf{Veto Authority}
  & \textbf{Documentation Required} \\
\midrule
\textbf{Critical}
  & Unacceptable risk per EU AI Act Art.~5; fundamental rights impact
  & Board + CPO/ChEO joint veto; automatic rejection
  & Full ethical impact assessment; board resolution; legal opinion \\[4pt]
\textbf{High}
  & High-risk per EU AI Act Annex~III; sensitive personal data; automated
    individual decision-making
  & AIGC majority + ChEO/CPO co-signature
  & Risk assessment; bias audit; data lineage; 6-month review \\[4pt]
\textbf{Elevated}
  & Moderate data sensitivity; limited automated decisions; new vendor AI
  & AIGC Chair (ChEO/CAIO) approval
  & Standard AIP form; legal review; 12-month monitoring plan \\[4pt]
\textbf{Standard}
  & Low sensitivity; human-in-the-loop; internal tooling
  & AIGC Secretariat approval
  & Abbreviated AIP; data classification confirmation \\
\bottomrule
\end{tabularx}
\end{table}

\subsubsection{Illustrative Application: Financial Services}
\label{ssec:finserv}

Financial services is among the most advanced sectors in AI governance
maturity, making it a rigorous and appropriate test case for the CAIGA. The
sector operates under overlapping regulatory frameworks, including Basel~III, GDPR,
MiFID~II, and, most directly, the UK Prudential Regulation Authority's
Supervisory Statement SS1/23 on Model Risk Management~\cite{PRASS123} and the European
Banking Authority's guidelines on internal governance~\cite{EBA2021}, each of which mandates
explicit model risk identification, cross-functional oversight, and
documented decision rationale. These requirements closely parallel the AIGC
workflow, providing a degree of empirical anchoring: banks and insurers that
have already implemented model risk committees (MRCs) in response to these
regulations are, in practice, operating a domain-specific precursor to the
CAIGA structure~\cite{Hadley2024}.

Consider a retail bank deploying an automated credit-scoring model
incorporating alternative data sources (social media signals, device metadata).
Under the CAIGA:

Risk Triage: The system is classified High Risk: it falls
within EU AI Act Annex~III (AI in access to essential services), processes
sensitive personal data, and makes automated decisions materially affecting
individuals' financial access~\cite{EUAIAct2024}.

Committee Review: Legal confirms GDPR Article~22 obligations (right
to explanation) and PRA SS1/23 model validation requirements. Cybersecurity
identifies data poisoning risks in the alternative data pipeline. Data Science
flags statistically significant demographic bias in social media signal
features. Business confirms commercial rationale and customer disclosure
requirements. Audit \& Risk documents the complete review trail.

Decision: Approved with Conditions: alternative data
features require a pre-deployment bias audit; a formal customer disclosure
statement is mandated; quarterly reporting to the CPO and ChEO for six months.

Outcome: The process surfaces demographic bias and regulatory
exposure that ad-hoc governance would likely have missed, illustrating how
the CAIGA converts ethical principles into concrete, auditable practice
consistent with the operationalization model proposed by M\"{a}ntym\"{a}ki
et al.~\cite{Mantymaki2022}. The alignment with existing MRC practice in
regulated financial institutions demonstrates that the committee structure is
not merely theoretical but is achievable within current organizational and
regulatory norms.

\section{Regulatory Alignment, Limitations, and Conclusion}
\label{sec:discussion_conclusion}

\subsection{Regulatory Alignment}
\label{ssec:regulatory_alignment}

The EU AI Act (Regulation 2024/1689), which came into force in August 2024,
represents the world's first comprehensive horizontal legal framework for AI
regulation~\cite{EUAIAct2024}. Its four-tier, risk-based classification
model, which includes Unacceptable, High, Limited, and Minimal risk categories, maps directly onto 
the CAIGA's decision matrix. Specifically, AI systems threatening fundamental
rights are considered an unacceptable risk under Article~5 and are prohibited outright, 
with these prohibitions becoming enforceable in February 2025. In the CAIGA, these map 
to the Critical tier, triggering automatic AIGC rejection with board ratification required. 
Furthermore, systems used in employment, essential services, education, and law enforcement 
are classified as high risk under Annex~III, facing stringent obligations that include mandatory
risk management, data governance, transparency, human oversight, and conformity
assessment. These align with the CAIGA's High tier, requiring AIGC
majority approval, formal bias audit, data lineage documentation, and a
structured post-deployment monitoring schedule. Finally, lower-risk systems facing lighter transparency
obligations under the limited and minimal risk categories correspond to the CAIGA's Elevated and
Standard tiers with proportionally lighter review requirements. Organizations implementing the CAIGA 
will be substantially positioned for EU AI Act compliance, since the committee's four-tier risk gate 
mirrors the Act's classification logic and provides the named accountable roles, documented
decision rationale, and audit trails that the Act requires of deployers. Beyond
the EU, the CAIGA maps onto the NIST AI Risk Management Framework (AI~RMF)~\cite{NIST2023}, as
detailed in Table~\ref{tab:nist_mapping}.

\begin{table}[!ht]
\caption{Alignment of CAIGA with NIST AI Risk Management Framework (AI RMF)}
\label{tab:nist_mapping}
\small
\centering
\begin{tabularx}{\linewidth}{@{} p{2.8cm} X @{}}
\toprule
\textbf{NIST AI RMF Function} & \textbf{CAIGA Component / Action} \\
\midrule
\textbf{Govern} & Executive AI Leadership (CPO \& ChEO) providing strategic
  oversight and accountability. \\[3pt]
\textbf{Map} & AIGC Risk Triage and AIP submission process identifying
  context and risks. \\[3pt]
\textbf{Measure} & AIGC cross-functional review, including bias audits and
  cybersecurity assessments. \\[3pt]
\textbf{Manage} & Post-deployment monitoring schedule, incident reporting,
  and re-review triggers. \\
\bottomrule
\end{tabularx}
\end{table}

This dual alignment positions the CAIGA as a governance architecture supporting compliance across multiple
jurisdictions, which is a critical advantage for multinational corporations.

\subsection{Limitations}
\label{ssec:limitations}

While the CAIGA provides a robust, structured approach to enterprise AI
governance, several limitations must be acknowledged. First, the framework
assumes the organizational scale of large enterprises with mature, distinct
functional departments (e.g., separate legal, cybersecurity, and data science
teams). Its scalability to small and medium-sized enterprises (SMEs), which
often consolidate these roles, remains empirically untested and may require a
condensed variant of the committee structure. Second, the current model lacks
a formalized disagreement resolution mechanism; future iterations must define
explicit voting rules or tie-breaking protocols for the AIGC when consensus
cannot be reached. Finally, the framework's risk classification and oversight
requirements are specifically calibrated to the EU AI Act and the NIST AI RMF.
Its jurisdictional scope and applicability in regions with divergent legal
definitions, cultural norms, or risk tolerances necessitate further evaluation.

\subsection{Conclusion}
\label{ssec:conclusion}

This paper has proposed the Corporate AI Governance Architecture~(CAIGA), a
structured three-tier framework that addresses the persistent gap between
high-level AI ethics principles and their concrete implementation in corporate
practice. By integrating specialized executive leadership, a cross-functional
AI Governance Committee with a formal go/no-go decision gate, and a
risk-tiered decision matrix aligned with the EU AI Act and NIST AI~RMF, the
CAIGA provides organizations with a replicable, auditable, and
regulation-ready governance infrastructure. Application to the financial services 
sector, anchored to documented model risk management practice under PRA SS1/23, 
illustrates how the framework surfaces risks that ad-hoc governance processes 
would likely miss.

Future research should empirically validate the CAIGA through case studies
across multiple sectors, examine the organizational conditions moderating its
effectiveness, and explore how the framework should evolve as agentic and
general-purpose AI systems challenge existing risk classification logic.

\section*{Acknowledgements}
The authors acknowledge the use of generative AI tools to improve the
readability and grammatical flow of this manuscript. The authors have reviewed
and edited the content and take full responsibility for the final text.

\end{document}